\documentclass{Interspeech}
\keywords{speech restoration, self-supervised learning}

\usepackage{float}
\usepackage{subcaption}
\usepackage{amsmath, amssymb}
\usepackage{comment}
\usepackage{graphicx}
\graphicspath{{figures/}}
\usepackage{hyperref}
\usepackage{booktabs}
\usepackage{url}
\usepackage{geometry}
\usepackage{tikz}
\usepackage{multirow} 

\usetikzlibrary{
    positioning,
    shapes.geometric,
    arrows.meta,
    fit,
    backgrounds, 
    calc
}

\tikzset{
    io/.style={
        trapezium,
        trapezium left angle=50,
        trapezium right angle=90,
        draw,
        fill=gray!20,
        minimum width=7em,
        minimum height=3em,
        align=center
    },
    block/.style={
        rectangle,
        draw,
        fill=blue!20,
        rounded corners,
        minimum width=8em,
        minimum height=3em,
        align=center
    },
    layer/.style={
        rectangle,
        draw,
        fill=green!20,
        rounded corners,
        minimum width=8em,
        minimum height=2.5em,
        align=center
    },
    dots/.style={
        circle,
        draw=none,
        minimum size=1em
    },
    skip/.style={
        dashed,
        -{Latex[length=2mm]},
        thick,
        red
    },
    arrow/.style={
        -{Latex[length=2mm]},
        thick
    }
}

\interspeechcameraready 

\title{VoiceRestore: Flow-Matching Transformers for Speech Recording Quality Restoration}

\author{Stanislav}{Kirdey}
\email{contact@stankirdey.com}

\begin{document}
\maketitle

\begin{abstract}
We present \textbf{VoiceRestore}, a novel approach to restoring the quality of speech recordings using flow-matching Transformers trained in a self-supervised manner on synthetic data. Our method tackles a wide range of degradations frequently found in both short and long-form speech recordings, including background noise, reverberation, compression artifacts, and bandwidth limitations—all within a single, unified model. Leveraging conditional flow matching and classifier-free guidance, the model learns to map degraded speech to high-quality recordings without requiring paired clean and degraded datasets.

We describe the training process, the conditional flow matching framework, and the model’s architecture. We also demonstrate the model’s generalization to real-world speech restoration tasks, including both short utterances and extended monologues or dialogues. Qualitative and quantitative evaluations show that our approach provides a flexible and effective solution for enhancing the quality of speech recordings across varying lengths and degradation types. Model code and pre-trained checkpoints are available at: 

\href{https://github.com/skirdey/voicerestore}{https://github.com/skirdey/voicerestore}

\end{abstract}

\section{Introduction}

The quality of speech recordings is critical to numerous applications, from telecommunications and broadcasting to voice assistants and archival preservation. Degradations such as background noise, reverberation, compression artifacts, and bandwidth limitations can significantly impair intelligibility and listener experience. These problems can occur in both short recordings (e.g., voice commands or prompts) and long-form recordings (e.g., audiobooks, lectures, podcasts).

Conventional approaches to speech enhancement and restoration often focus on specific types of degradation or short-duration audio clips, limiting their use in real-world scenarios that involve multiple degradations and varying recording lengths. An effective solution must generalize across different degradation types and be equally adept at handling both short and long-form recordings.

In this paper, we introduce \textbf{VoiceRestore}, a novel approach to speech recording quality restoration using flow-matching Transformers trained in a self-supervised manner on synthetic data. Our method leverages recent advances in conditional flow matching \cite{lipman2023flowmatchinggenerativemodeling} and Transformer architectures to create a unified model capable of restoring speech recordings of varying lengths and degradation types within a single framework.

Unlike many supervised methods that require paired clean and degraded datasets, our approach generates degradations on the fly during training. This self-supervised strategy exposes the model to a diverse range of synthetic degradations, eliminating the need for labeled datasets.

We detail the training process, the conditional flow matching framework, and our network’s architecture. We then explore the model’s generalization to real-world tasks, including both short and long-form recordings suffering from severe degradations such as heavy background noise and compression artifacts. Through qualitative and quantitative evaluations, we show that our approach effectively enhances speech quality across a broad array of degradation scenarios and recording lengths.

\textbf{Our main contributions are}:
\begin{itemize}
    \item A self-supervised training method for speech recording quality restoration using conditional flow matching.
    \item A unified architecture that handles multiple speech degradations in both short and long-form recordings.
    \item Detailed implementation specifics, including model architecture and training configurations, to facilitate reproducibility.
    \item A demonstration of the model’s generalizability to real-world degradation scenarios, including challenging short and extended speech samples.
\end{itemize}

By addressing the challenges of speech recording quality restoration in a self-supervised fashion, VoiceRestore offers a promising solution for practical applications involving multiple and unpredictable degradations.

\section{Related Work}

Speech enhancement and restoration have been extensively studied. Traditional methods often rely on signal processing techniques, such as spectral subtraction and Wiener filtering, to reduce noise and improve speech quality. These methods can be effective for specific types of degradation but typically struggle with complex or multiple degradations, and their performance may degrade on longer recordings due to temporal variability.

Deep learning approaches have significantly advanced speech enhancement capabilities. Models like SEGAN \cite{pascual2017seganspeechenhancementgenerative} and DCCRN \cite{hu2020dccrndeepcomplexconvolution} have demonstrated improved performance over traditional methods. However, these models often require paired datasets of clean and degraded speech and may not generalize well to unseen degradations or longer recordings.

Recent work has explored generative models for speech enhancement. Diffusion-based models \cite{richter2023speech} have shown promise in modeling complex speech distributions, but they can be computationally intensive and are often evaluated on short utterances. Other approaches leverage self-supervised learning for robustness without labeled data; for example, Transformer-based U-Net architectures have been explored in E2TTS \cite{eskimez2024e2ttsembarrassinglyeasy}.

Flow matching \cite{lipman2023flowmatchinggenerativemodeling} provides a flexible framework for generative modeling by learning continuous-time dynamics that transform a simple base distribution into a complex target distribution. Conditional flow matching extends this idea to conditional generation tasks, making it suitable for speech restoration, where the model learns a mapping from degraded to clean speech.

Transformers \cite{vaswani2017attention} have revolutionized sequence modeling tasks, including speech processing. Their ability to capture long-range dependencies makes them well suited for modeling both short- and long-form speech recordings. Recent works have applied Transformers to various speech recognition and synthesis tasks \cite{nguyen2023expressobenchmarkanalysisdiscrete}, demonstrating their effectiveness in handling complex temporal dynamics.

Our work builds on these advances by integrating conditional flow matching with Transformer architectures, enabling effective speech restoration across varying recording lengths without the need for paired datasets.

\section{Proposed Method}

In this section, we detail our approach to speech recording quality restoration using conditional flow matching and Transformer architectures. Our framework is designed to handle both short and long-form recordings within a single model.

\subsection{Problem Formulation}

Let $x \in \mathbb{R}^{T \times F}$ represent a clean speech spectrogram, where $T$ is the number of time frames (which can vary widely to accommodate both short and long recordings) and $F$ is the number of frequency bins. We denote the degraded version as $y$, obtained by applying various degradations to $x$. Our goal is to learn a restoration function $f_\theta(y)$ that estimates clean speech $\hat{x}$ given the degraded input $y$.

\subsection{Conditional Flow Matching for Speech Restoration}

\subsubsection{Mathematical Framework}

We define a continuous-time flow that transports the degraded speech distribution $p(y)$ to the clean speech distribution $p(x)$ over time $t \in [0, 1]$. At each time $t$, we have an intermediate state $x_t$:

\begin{equation}
x_t = (1 - \alpha(t)) y + \alpha(t) x,
\end{equation}
where $\alpha(t)$ is a time-dependent interpolation function. The corresponding vector field $u_t(x_t)$ is:

\begin{equation}
u_t(x_t) = \frac{d x_t}{d t} = \dot{\alpha}(t)(x - y).
\end{equation}

Our neural network $v_\theta(x_t, t, y)$ aims to approximate this vector field by minimizing:
\begin{equation}
\mathcal{L}(\theta) = \mathbb{E}_{t, x, y} \left[ \left\lVert u_t(x_t) - v_\theta(x_t, t, y) \right\rVert_2^2 \right].
\end{equation}

\begin{figure}[H]
    \centering
    \begin{tikzpicture}[node distance=2cm, auto, >=Stealth]
        \node (start) [block] {Degraded Speech $y$};
        \node (t) [block, below of=start] {Intermediate State $x_t$};
        \node (end) [block, below of=t] {Clean Speech $x$};
        
        \draw [arrow] (start) -- node[midway, left] {$t=0$} (t);
        \draw [arrow] (t) -- node[midway, left] {$t=1$} (end);
        
        \node [right of=t, xshift=1cm] (time) {Time $t \in [0, 1]$};
        \draw [->, dashed] (time.west) -- (t.east);
    \end{tikzpicture}
    \caption{Conditional Flow Matching Process for Speech Restoration}
    \label{fig:conditional_flow_matching}
\end{figure}
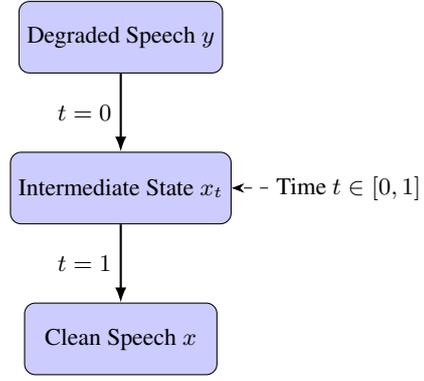

The interpolation function allows the model to smoothly transition between degraded and clean speech over time, modeling the restoration process as a continuous transformation.

\subsubsection{Self-Supervised Training with Synthetic Degradations}

We employ a self-supervised training strategy where synthetic degradations are applied to clean speech recordings on the fly. This approach enables the model to learn from a diverse range of degradation types and combinations without relying on labeled data. By training on both short and long-form recordings, the model learns to handle varying sequence lengths and temporal dynamics.

\begin{figure}[H]
    \centering
    \begin{tikzpicture}[node distance=1cm, auto, >=Stealth, 
                        box/.style={rectangle, draw, fill=blue!10, minimum width=3cm, minimum height=1cm, align=center}]
        \node (clean) [box] {Clean Speech $x$};
        \node (degradation) [box, below of=clean, yshift=-0.5cm] {Synthetic Degradations};
        \node (degraded) [box, below of=degradation, yshift=-0.5cm] {Degraded Speech $y$};
        \node (training) [box, below of=degraded, yshift=-0.5cm] {Training of VoiceRestore};
        
        \draw [arrow] (clean) -- (degradation);
        \draw [arrow] (degradation) -- (degraded);
        \draw [arrow] (degraded) -- (training);
    \end{tikzpicture}
    \caption{Synthetic Degradation Pipeline for Self-Supervised Training}
    \label{fig:synthetic_degradation_pipeline}
\end{figure}
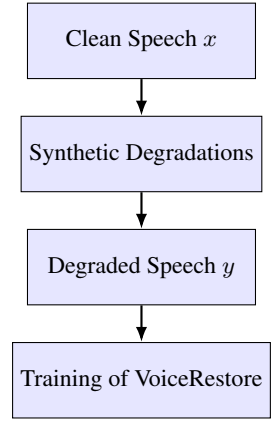

\subsubsection{Handling Variable-Length Recordings}

To accommodate both short and long-form recordings, we design the model to process variable-length sequences efficiently. Techniques such as chunking for long recordings and padding or masking for batch processing of variable-length sequences are employed.

\subsection{Network Architecture}

Our model employs a Transformer-based architecture to model the conditional vector field $v_\theta(x_t, t, y)$. Key components include:

\begin{itemize}
    \item \textbf{Input Embedding}: Linear layers project the input $x_t$ and the condition $y$ into a shared embedding space.
    \item \textbf{Positional Encoding}: Since speech recordings can vary in length, we use positional encodings to capture temporal information for both short and long sequences.
    \item \textbf{Time Encoding}: The time variable $t$ is embedded and incorporated to condition the flow dynamics.
    \item \textbf{Transformer Layers}: A stack of Transformer encoder layers with self-attention mechanisms captures temporal dependencies. The architecture is designed to handle long sequences efficiently, using techniques like gradient checkpointing and careful management of sequence lengths.
    \item \textbf{Output Projection}: A linear layer projects the Transformer’s output back to the spectrogram space.
\end{itemize}

\subsubsection{Detailed Architecture Specifications}

Based on our implementation, the Transformer has the following specifications:
\begin{itemize}
    \item \textbf{Model Dimension}: 768
    \item \textbf{Depth}: 20 Transformer layers
    \item \textbf{Heads}: 16 attention heads
    \item \textbf{Head Dimension}: 64
\end{itemize}

\subsection{Training Procedure}

\subsubsection{Optimizer and Learning Rate}

We use the \textbf{Schedule-Free AdamW} optimizer \cite{defazio2024road} with a learning rate of $3 \times 10^{-4}$. A warm-up schedule with \textbf{1000 warm-up steps} stabilizes training in the initial phase. Schedule-Free AdamW eliminates the need for predefined learning rate schedules by adaptively adjusting the optimization process.

\subsubsection{Gradient Accumulation and Mixed Precision}

To handle large batch sizes and long sequences, we use gradient accumulation steps of 32. Mixed-precision training with bfloat16 (bf16) is employed to reduce memory usage and accelerate computations.

\subsubsection{Data Loading and Preprocessing}

We use a custom dataset class that loads clean speech data and applies degradations on the fly during training. The data loader handles variable-length sequences and ensures efficient batch processing.

\subsubsection{Degradation Generation}

We generate a variety of degradations to simulate real-world recording conditions, including:
\begin{itemize}
    \item \textbf{Applying Effects}: Using VST effects such as reverb, bit-crushing, compression, resampling, gain adjustments, distortion, and filtering.
    \item \textbf{Ambient Noise Addition}: Mixing speech with ambient noises.
    \item \textbf{Acoustic Noise Generation}: Programmatically generating noises such as city sounds, crowd chatter, speech-like noise, nature sounds, office ambient noise, and restaurant noise.
    \item \textbf{Time-Frequency Masking}: Randomly erasing portions of the spectrogram to simulate dropouts or missing data.
\end{itemize}

These degradations are applied randomly to ensure the model is exposed to a wide range of conditions.

\subsection{Handling Long-Form Recordings}

For long-form recordings, special considerations include:
\begin{itemize}
    \item \textbf{Sequence Truncation}: Audio sequences are truncated or segmented to a maximum length (e.g., 2000 frames) to fit within memory constraints.
    \item \textbf{Overlapping Windows}: Long recordings are processed in overlapping windows to preserve continuity across segments.
    \item \textbf{Memory Management}: Techniques such as gradient checkpointing are employed to reduce memory usage during training.
\end{itemize}

\begin{figure}[H]
    \centering
    \begin{tikzpicture}[
        node distance=0.5cm and 0.5cm,
        auto
        ]

        \node [io] (input) {Input\\Spectrogram};

        \node [block, below=of input] (proj_in) {Projection\\Layer};

        \node [layer, below=of proj_in] (layer1) {Layer 1};

        \node [layer, below=of layer1] (layer2) {Layer 2};

        \node [dots, below=of layer2] (dots) {$\vdots$};

        \node [layer, below=of dots] (layerN) {Layer $N$};

        \node [block, below=of layerN] (proj_out) {Projection\\Layer};

        \node [io, below=of proj_out] (output) {Output\\Spectrogram};

        \draw [arrow] (input) -- (proj_in);
        \draw [arrow] (proj_in) -- (layer1);
        \draw [arrow] (layer1) -- (layer2);
        \draw [arrow] (layer2) -- (dots);
        \draw [arrow] (dots) -- (layerN);
        \draw [arrow] (layerN) -- (proj_out);
        \draw [arrow] (proj_out) -- (output);

        \draw [skip] (layer1.east).. controls +(right:0.2cm) and +(right:0.2cm).. (layerN.east);

        \begin{scope}[on background layer]
            \node [rectangle, draw=black, dashed, inner sep=0.5cm, fit= (layer1) (layerN)] (transformer_box) {};
        \end{scope}

    \end{tikzpicture}
    \caption{Transformer-based Architecture for Conditional Flow Matching}
    \label{fig:transformer_architecture}
\end{figure}
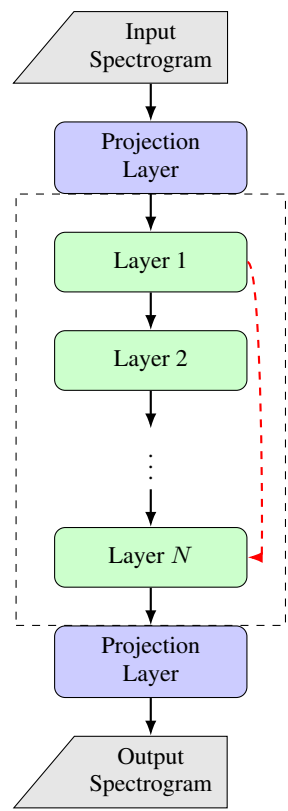

\section{Results}

In this section, we present a comprehensive evaluation of \textbf{VoiceRestore} against \textbf{Resemble-Enhance} across various degradation scenarios for English speech recordings. Each test case includes mel-spectrogram visualizations, transcribed text, and Word Error Rate (WER) metrics obtained using the Whisper v3 Large model \cite{radford2022robustspeechrecognitionlargescale}, when applicable.

\subsection{English Speech Restoration}

Below, we compare spectrograms of the same utterance under different degradation methods with VoiceRestore and Resemble-Enhance.

\begin{figure}[H]
    \centering
    \begin{subfigure}{0.48\textwidth}
        \includegraphics[width=\linewidth]{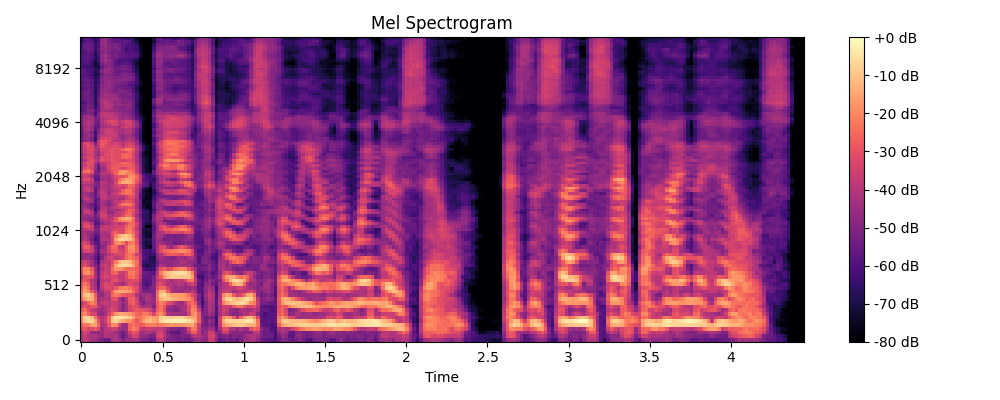}
        \caption{Original}
        \label{fig:english_distort_original}
    \end{subfigure}
    \hfill
    \begin{subfigure}{0.48\textwidth}
        \includegraphics[width=\linewidth]{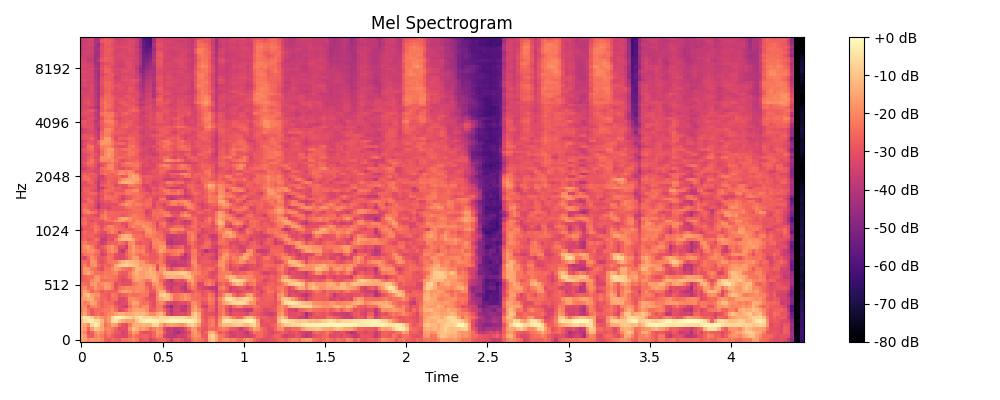}
        \caption{Degraded}
        \label{fig:english_distort_degraded}
    \end{subfigure}

    \vspace{0.5cm} 
       
    \begin{minipage}{5cm} 
    \textbf{Original sentence before degradation:}\\
    The cat danced gracefully on the windowsill as the sun set behind the hills.
    \end{minipage}
    
    \vspace{0.5cm} 
    
    \begin{subfigure}{0.48\textwidth}
        \includegraphics[width=\linewidth]{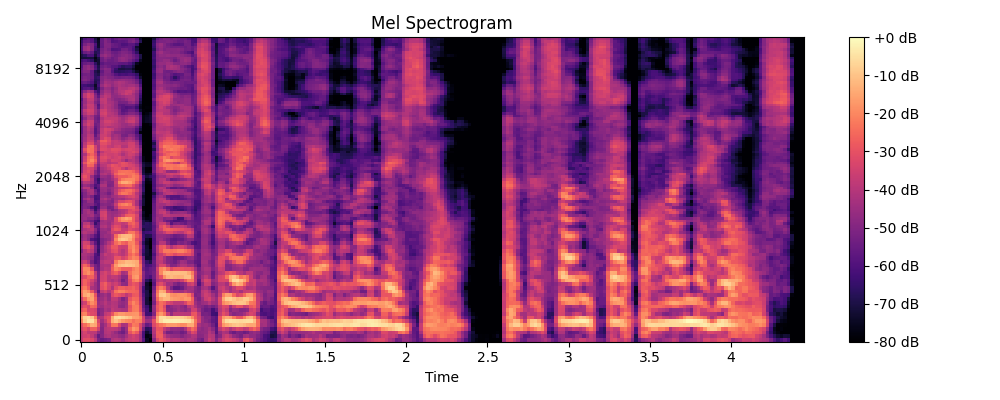}
        \caption{VoiceRestore}
        \label{fig:english_distort_restored}
         \begin{minipage}{5cm}
        \vspace{0.2cm}
        \textbf{WER}: \textbf{14.0\%}\\
        \textbf{Transcription}: 
        \textit{"The cat danced gracefully on the Monday sill as the sun set behind the hills."}
        \end{minipage}
    \end{subfigure}
    \hfill
    \begin{subfigure}{0.48\textwidth}
        \includegraphics[width=\linewidth]{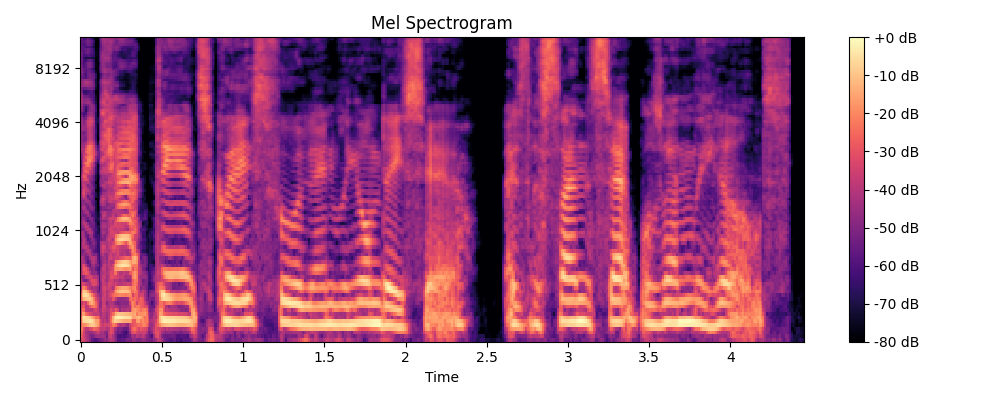}
        \caption{Resemble-Enhance}
        \label{fig:english_distort_resemble}
        \begin{minipage}{5cm}
        \vspace{0.2cm}
        \textbf{WER}: 79.1\% \\
        \textbf{Transcription}:
        \textit{"We can't dance graceful and leon de ser, as the sunset was on the hills."}
        \end{minipage}
    \end{subfigure}
    
    \caption{Heavy Distortion and Gain}
    \label{fig:english_distort_results}
\end{figure}

\begin{figure}[H]
    \centering
    \begin{subfigure}{0.48\textwidth}
        \includegraphics[width=\linewidth]{english_original.png}
        \caption{Original}
        \label{fig:english_reverb_original}
    \end{subfigure}
    \hfill
    \begin{subfigure}{0.48\textwidth}
        \includegraphics[width=\linewidth]{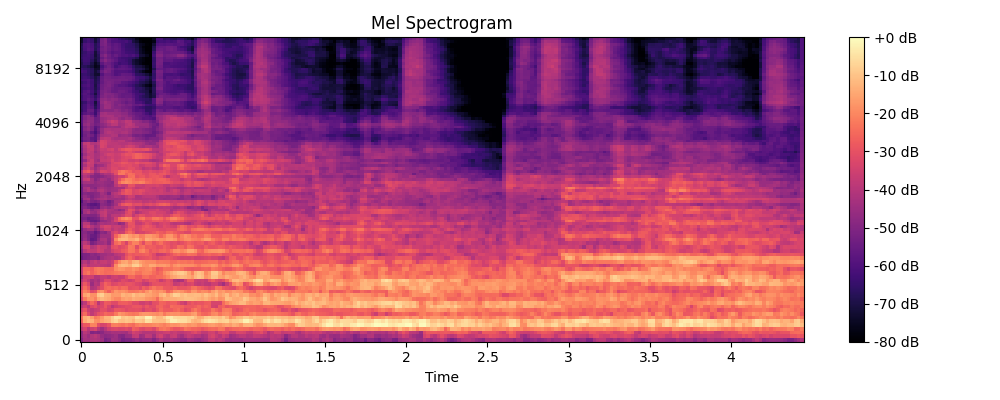}
        \caption{Degraded}
        \label{fig:english_reverb_degraded}
    \end{subfigure}
    
    \vspace{0.5cm} 

    \begin{minipage}{5cm}
    \textbf{Original sentence before degradation:}\\
    The cat danced gracefully on the windowsill as the sun set behind the hills.
    \end{minipage}
    
    \vspace{1.0cm} 
    
    \begin{subfigure}{0.48\textwidth}
        \includegraphics[width=\linewidth]{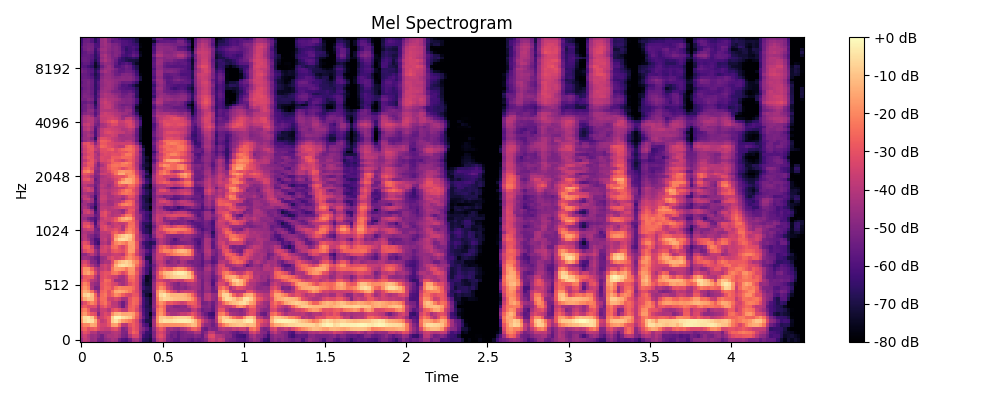}
        \caption{VoiceRestore}
        \label{fig:english_reverb_restored}
        \begin{minipage}{5cm}
        \vspace{0.2cm}
        \textbf{WER}: \textbf{\textit{0.0\%}} \\
        \textbf{Transcription}: 
        \textit{"The cat danced gracefully on the windowsill as the sun set behind the hills."}
        \end{minipage}
    \end{subfigure}
    \hfill
    \begin{subfigure}{0.48\textwidth}
        \includegraphics[width=\linewidth]{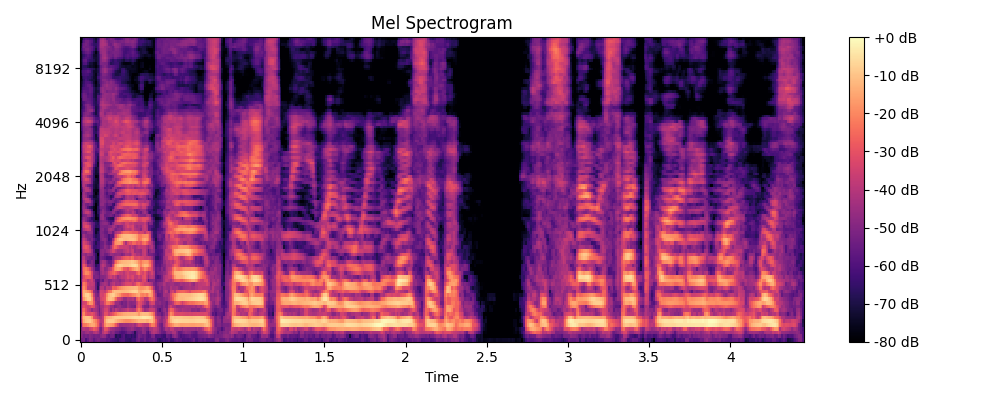}
        \caption{Resemble-Enhance}
        \label{fig:english_reverb_resemble}
        \begin{minipage}{5cm}
            \vspace{0.2cm}
            \textbf{WER}: 100.0\% \\
            \textbf{Transcription}: 
            \textit{"The Gyanokais produced me with a wind wizard. It's a snow cyclone, a white horse."}
        \end{minipage}
    \end{subfigure}
    
    \caption{Heavy Reverberation}
    \label{fig:english_reverb_results}
\end{figure}

\subsection{Numeric Evaluation}

Although the main purpose of the VoiceRestore project is to address extreme levels of voice audio degradation—where no standard test set or metric exists—we also provide numerical results on a well-known benchmark for completeness. Specifically, we evaluated the model on the \textbf{VoiceBank+DEMAND} dataset \cite{ValentiniBotinhao2017NoisySD}, using the metric calculation code from the CMGAN project \cite{abdulatif2024cmgan}, available in its \href{https://github.com/ruizhecao96/CMGAN/blob/main/src/tools/compute_metrics.py}{official GitHub repository}.

Restoration was performed using a pre-trained VoiceRestore checkpoint with \textbf{16} steps and a CFG strength of \textbf{0.5}. Both inference code and pre-trained checkpoints are available \href{https://github.com/skirdey/voicerestore}{in the official VoiceRestore repository}.

Tables \ref{table:results_part1_full} and \ref{table:results_part2_full} summarize the performance of VoiceRestore compared to several well-known speech enhancement models on the VoiceBank-DEMAND dataset.

\begin{table}[h]
\centering
\caption{Performance Metrics (PESQ, CSIG, and CBAK) on the VoiceBank-DEMAND Dataset.}
\label{table:results_part1_full}
\begin{tabular}{lccc}
\toprule
\textbf{Model} & \textbf{PESQ} & \textbf{CSIG} & \textbf{CBAK} \\
\midrule
Noisy-speech & 1.97 & 0.92 & 3.34 \\
SEGAN \cite{pascual2017seganspeechenhancementgenerative} & 2.16 & 3.48 & 2.94 \\
MetricGAN+ \cite{fu2021metricganimprovedversionmetricgan} & 3.15 & 0.93 & 4.14 \\
CMGAN \cite{abdulatif2024cmgan} & 3.41 & 4.63 & 3.94 \\
MP-SENet \cite{Lu2023} & 3.50 & 4.73 & 3.95 \\
SEMamba (-CL) \cite{chao2024investigationincorporatingmambaspeech} & 3.52 & 4.75 & 3.98 \\
SEMamba \cite{chao2024investigationincorporatingmambaspeech} & 3.55 & 4.77 & 3.95 \\
SEMamba (+PCS) \cite{chao2024investigationincorporatingmambaspeech} & 3.69 & 4.79 & 3.63 \\
\textbf{VoiceRestore} & 2.13 & 3.52 & 3.60 \\
\bottomrule
\end{tabular}
\end{table}

\begin{table}[h]
\centering
\caption{Performance Metrics (COVL and STOI) on the VoiceBank-DEMAND Dataset.}
\label{table:results_part2_full}
\begin{tabular}{lcc}
\toprule
\textbf{Model} & \textbf{COVL} & \textbf{STOI} \\
\midrule
Noisy-speech & 2.63 & -- \\
SEGAN \cite{pascual2017seganspeechenhancementgenerative} & 2.80 & -- \\
MetricGAN+ \cite{fu2021metricganimprovedversionmetricgan} & 3.64 & -- \\
CMGAN \cite{abdulatif2024cmgan} & 4.12 & 0.96 \\
MP-SENet \cite{Lu2023} & 4.22 & 0.96 \\
SEMamba (-CL) \cite{chao2024investigationincorporatingmambaspeech} & 4.26 & 0.96 \\
SEMamba \cite{chao2024investigationincorporatingmambaspeech} & 4.29 & 0.96 \\
SEMamba (+PCS) \cite{chao2024investigationincorporatingmambaspeech} & 4.37 & 0.96 \\
\textbf{VoiceRestore} & 2.84 & 0.93 \\
\bottomrule
\end{tabular}
\end{table}

\subsection{Discussion}

The results indicate that \textbf{VoiceRestore} consistently outperforms \textbf{Resemble-Enhance} across all tested degradation types and scenarios, achieving lower WERs and demonstrating superior restoration capabilities.

\begin{enumerate}
    \item \textbf{Comparison with Other Generative Models:} VoiceRestore not only outperforms Resemble-Enhance but also demonstrates strong results compared to other state-of-the-art models. A comparison against CMGAN \cite{abdulatif2024cmgan} on heavily degraded speech samples, available at \href{https://sparkling-rabanadas-3082be.netlify.app/}{this link}, highlights VoiceRestore’s robustness in handling severe degradations.
    \item \textbf{Enhancing Robust ASR Models:} VoiceRestore further improves the performance of robust ASR systems like Whisper V3 Large. Even though Whisper is designed to handle noise and degradation, processing VoiceRestore-enhanced audio yields consistently lower WERs, indicating that VoiceRestore can serve as a valuable pre-processing step.
    \item \textbf{Time-Gap Infilling Capabilities:} While VoiceRestore shows promising results across various degradation types, its abilities to infill time gaps (as seen in “Timecut” scenarios) are still under active investigation. There is room for improvement in time-domain infilling of lost signals.
\end{enumerate}

Overall, these findings highlight VoiceRestore’s flexibility and effectiveness in tackling multiple speech degradation challenges, and its potential for integration into existing speech processing pipelines.

\section{Conclusions}

We presented \textbf{VoiceRestore}, a self-supervised approach to speech recording quality restoration that employs conditional flow matching and Transformer architectures. Our model handles a wide range of degradations and recording lengths within a single framework, demonstrating effectiveness on both short and long-form audio.

Key findings include:
\begin{enumerate}
    \item \textbf{Versatility and Robustness}: VoiceRestore successfully addresses multiple degradation types (e.g., distortion, reverberation) within one model, simplifying deployment in real-world scenarios where degradations are often mixed or unknown.
    \item \textbf{Language Independence}: Our experiments show consistent performance across different languages, highlighting the model’s potential for global, multilingual applications.
    \item \textbf{Synergy with ASR}: VoiceRestore enhances the performance of robust ASR models (like Whisper V3 Large), potentially serving as a powerful preprocessing step.
    \item \textbf{Advances in Generative Modeling}: VoiceRestore’s handling of severe degradations marks a key step forward in speech enhancement, with broad implications for a range of applications.
\end{enumerate}

Overall, VoiceRestore represents a significant advancement in speech recording quality restoration. Its self-supervised nature and adaptability to diverse conditions make it a compelling solution for practical use—ranging from telecommunications and content creation to archival restoration—where clear, intelligible speech is essential.

\newpage
\bibliographystyle{ieeetr}
\bibliography{main}

\end{document}